\documentclass[twocolumn]{aastex63}
\usepackage{natbib}

\usepackage{hyperref}

\graphicspath{{./}{figures/}}
\usepackage{graphics,graphicx}
\usepackage{footnote}

\newcommand{\uat}[2]{\href{http://astrothesaurus.org/uat/#2}{#1 (#2)}}
\newcommand{\shortname}{PJ231$-$20}
\newcommand{\longname}{PSO J231.6576$-$20.8335}
\newcommand\luv{$L_{2500\, \textup{\footnotesize \AA}}$}
\newcommand\fuv{$F_{2500\, \textup{\footnotesize \AA}}$}

\received{May 26, 2020}
\revised{July 14, 2020}
\accepted{July 18, 2020}
\submitjournal{The Astrophysical Journal}

\shorttitle{{\it Chandra} Observations of PJ231$-$20}
\shortauthors{Connor et al.}
\graphicspath{{./}{figures/}}

\begin{document}

\title{X-ray Observations of a [\ion{C}{2}]-bright, $z=6.59$ Quasar/Companion System}

\correspondingauthor{Thomas Connor}
\email{thomas.p.connor@jpl.nasa.gov}

\author[0000-0002-7898-7664]{Thomas Connor}
\affiliation{Jet Propulsion Laboratory, California Institute of Technology, 4800 Oak Grove Drive, Pasadena, CA 91109, USA}
\affiliation{The Observatories of the Carnegie Institution for Science, 813 Santa Barbara St., Pasadena, CA 91101, USA}

\author[0000-0002-2931-7824]{Eduardo Ba\~nados}
\affiliation{Max Planck Institute for Astronomy, K\"onigstuhl 17, 69117 Heidelberg, Germany}
\affiliation{The Observatories of the Carnegie Institution for Science, 813 Santa Barbara St., Pasadena, CA 91101, USA}

\author[0000-0002-5941-5214]{Chiara Mazzucchelli}
\affiliation{European Southern Observatory, Alonso de Cordova 3107, Vitacura, Region Metropolitana, Chile}

\author[0000-0003-2686-9241]{Daniel Stern}
\affiliation{Jet Propulsion Laboratory, California Institute of Technology, 4800 Oak Grove Drive, Pasadena, CA 91109, USA}

\author[0000-0002-2662-8803]{Roberto Decarli}
 \affiliation{INAF --- Osservatorio di Astrofisica e Scienza dello Spazio, via Gobetti 93/3, I-40129, Bologna, Italy}

\author[0000-0003-3310-0131]{Xiaohui Fan}
\affil{Steward Observatory, University of Arizona, 933 N Cherry Ave, Tucson, AZ, USA}

\author[0000-0002-6822-2254]{Emanuele Paolo Farina}
\affiliation{Max Planck Institut f\"ur Astrophysik, Karl--Schwarzschild--Stra{\ss}e 1, D-85748, Garching bei M\"unchen, Germany}

\author[0000-0003-0083-1157]{Elisabeta Lusso}
\affiliation{Dipartimento di Fisica e Astronomia, Universit\`a di Firenze, via G. Sansone 1, I-50019 Sesto Fiorentino, Firenze, Italy}
\affiliation{INAF -- Osservatorio Astrofisico di Arcetri, Largo Enrico Fermi 5, I-50125 Firenze, Italy}

\author[0000-0002-9838-8191]{Marcel Neeleman}
\affiliation{Max Planck Institute for Astronomy, K\"onigstuhl 17, 69117 Heidelberg, Germany}

\author[0000-0003-4793-7880]{Fabian Walter}
\affil{Max Planck Institute for Astronomy, K\"onigstuhl 17, 69117 Heidelberg, Germany}

\begin{abstract}

We present deep {\it Chandra} observations of \longname, a quasar at redshift $z=6.59$ with a nearby (${\sim}8$ proper kpc) companion galaxy. ALMA observed both the quasar and companion to be bright in [\ion{C}{2}], and the system has significant extended Ly$\alpha$ emission around the quasar, suggesting that a galaxy merger is ongoing. Unlike previous studies of two similar systems, and despite observing the system with {\it Chandra} for 140 ks, we do not detect the companion in X-rays. The quasar itself is detected, but only $13.3^{+4.8}_{-3.7}$ net counts are observed. From a basic spectral analysis, the X-ray spectrum of the quasar is soft (hardness ratio of $\mathcal{HR} = -0.60_{-0.27}^{+0.17}$, power-law index of $\Gamma=2.6^{+1.0}_{-0.9}$), which results in a rest-frame X-ray luminosity comparable to other bright quasars ($L_{2-10} = 1.09^{+2.20}_{-0.70}\times 10^{45}\ \textrm{erg}\ \textrm{s}^{-1}$) despite the faint observed X-ray flux. We highlight two possible interpretations of this result: the quasar has a steep value of $\Gamma$ -- potentially related to observed ongoing Eddington accretion -- thereby pushing much of the emission out of our observed band, or the quasar has a more normal spectrum ($\Gamma{\sim}2$) but is therefore less X-ray luminous  ($L_{2-10} \sim 0.6 \times 10^{45}\ \textrm{ erg}\ \textrm{ s}^{-1}$).

\end{abstract}


\section{Introduction} \label{sec:intro}

In the first billion years of the universe ($z \gtrsim 5.7$), supermassive black holes (SMBHs) grew from initial seeds to masses of over $10^9\ {\rm M}_{\odot}$ \citep[e.g.,][]{2019ApJ...873...35S}; observed as quasars, the significant early growth of these objects remains an outstanding challenge for cosmology \citep[e.g.][]{2019ConPh..60..111S}. 
Driven by deep infrared surveys and comprehensive spectroscopic followup campaigns, the number of quasars known in this redshift regime has risen drastically in the past decade \citep[e.g.,][]{2014AJ....148...14B,2016ApJS..227...11B,2018Natur.553..473B, 2015ApJ...801L..11V, 2016ApJ...833..222J, 2017ApJ...849...91M, 2017MNRAS.468.4702R,2019MNRAS.487.1874R, 2019ApJ...872L...2M, 2019AJ....157..236Y, 2019ApJ...884...30W}, providing new opportunities to investigate black hole growth mechanisms. X-ray observations, which offer the best view of the inner accretion region of the active galactic nuclei (AGN) powering quasars \citep{2016AN....337..375F}, are of particular importance for understanding how SMBHs are able to grow so quickly. 
\begin{figure*}
\begin{center}
\includegraphics{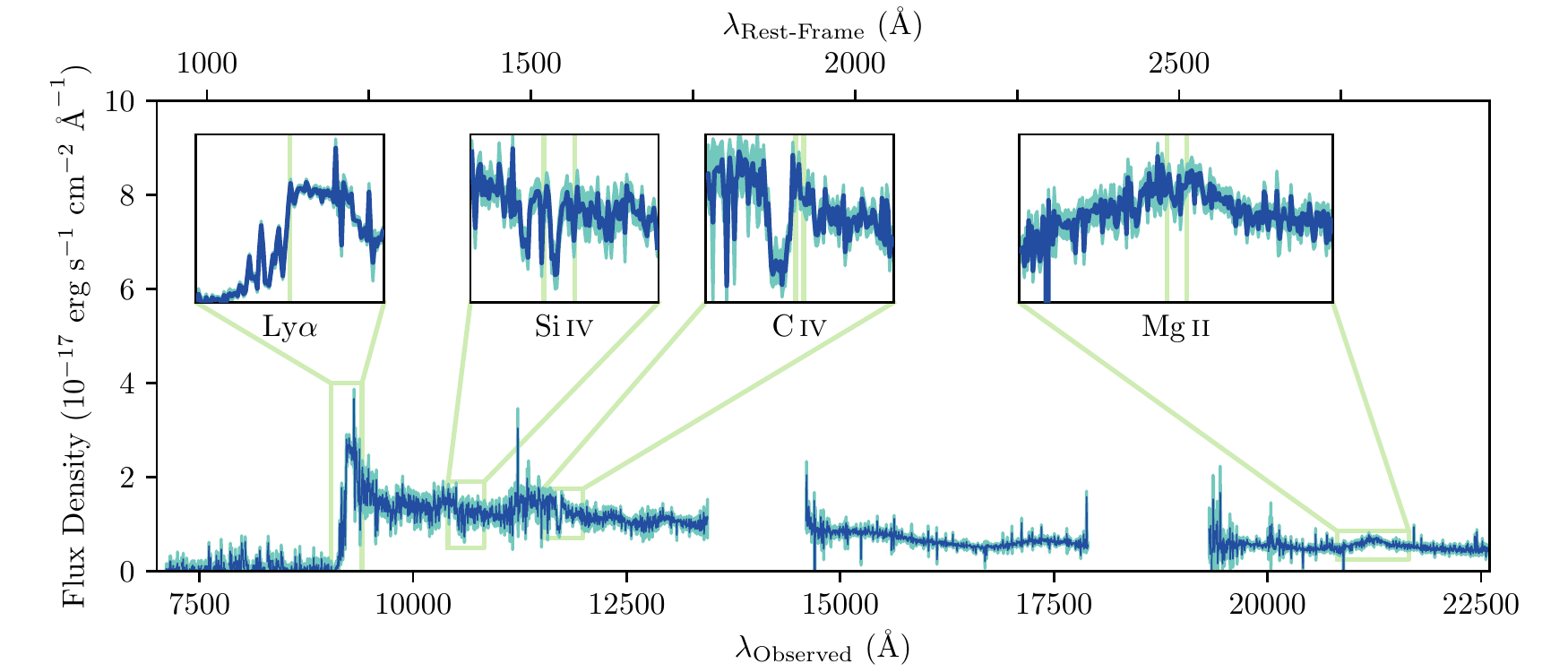}
\end{center}
\caption{Near-infrared spectrum of \shortname\ (blue) adapted from \citet{2017ApJ...849...91M}, with $1\sigma$ errors shown in teal. Per that work, this spectrum is flux calibrated to the $J$-band and corrected for the intrinsic reddening of the Milky Way. Four spectral transitions of note are detailed in cutouts, with the wavelength of the associated lines at the adopted redshift of the quasar indicated by the vertical green lines. The spectrum shows high velocity troughs blueshifted from the high excitation lines (\ion{Si}{4} and \ion{C}{4}), and we therefore classify \shortname\ as a HiBAL.}\label{fig:nir_spec}
\end{figure*}

While the population of quasars at the highest redshifts has long been speculated to reside in overdense regions \citep{2014MNRAS.439.2146C}, in recent years a number of works using the Atacama Large Millimeter/submillimeter Array (ALMA) have revealed that a number of these quasars have [\ion{C}{2}]-bright companions (\citealt{2017Natur.545..457D}, \citeyear{2018ApJ...854...97D}; \citealt{2017ApJ...850..108W}; \citealt{2019ApJ...882...10N}; \citealt{2019ApJ...874L..30V}; see also \citealt{2017ApJ...848...78F}; \citealt{2018ApJ...863L..29D}). These gas-rich mergers could potentially seed the rapid growth required to explain the early population of SMBHs \citep{2008ApJS..175..356H}. In addition, if companion galaxies also host AGN, the resultant SMBH mergers \citep{2005ApJ...630..152E}, which are invisible to current gravitational wave observatories \citep[e.g.][]{2016MNRAS.463..870S}, could produce rapid super-Eddington growth; this stochastic growth would, in turn, reduce the need for sustained super-Eddington accretion.

So far, companion galaxies of $z>6$ quasars are mostly unexplored in X-ray observations, with two exceptions. \citet{2019A&A...628L...6V} studied PSO J167.6415$-$13.4960, a $z=6.52$ quasar with a companion detected in both [\ion{C}{2}] and rest-frame UV emission ${\sim}5\,\textrm{pkpc}$ ($0\farcs9$) away \citep{2017ApJ...850..108W, 2019ApJ...881..163M}. In a 59 ks {\it Chandra} observation, \citet{2019A&A...628L...6V} detected three hard energy photons at the location of the system -- a significant ($P=0.9996$) detection implying a heavily obscured quasar, but one which could not be clearly assigned to either the quasar or its companion. Recently, \citet{2019ApJ...887..171C} reported on a 150 ks {\it Chandra} observation of PSO~J308.0416$-$21.2339, a $z=6.23$ quasar with a [\ion{C}{2}]- and UV-bright companion visible on both sides of the quasar, indicative of an ongoing merger \citep{2017Natur.545..457D, 2019ApJ...880..157D}. \citet{2019ApJ...887..171C} detected three hard energy photons at the brightest knot of UV emission indicative of a dual AGN, but, due to the longer exposure time, the detection was less significant ($P=0.979$). With no other deep X-ray observations of high-redshift quasar companions reported, further observations are needed to constrain the roles of companions in SMBH growth.

In this work, we focus on and present new deep {\it Chandra} observations of \longname\ (hereafter \shortname), a $z=6.5864\pm0.0005$ quasar first discovered by \citet{2017ApJ...849...91M}. \shortname's systemic redshift was measured by \citet{2018ApJ...854...97D} from [\ion{C}{2}] emission observed with ALMA. In those same ALMA observations, a companion galaxy was detected $8.4\pm0.6\,\textrm{pkpc}$ (${\sim}1\farcs5$) away at a relative velocity of $137\,\textrm{km}\,\textrm{s}^{-1}$ \citep[$z=6.5900$,][]{2017Natur.545..457D}. Deeper and higher-resolution ALMA observations presented by \citet{2019ApJ...882...10N} distinctly resolved both objects\footnote{\citet{2019ApJ...882...10N} also report a second companion, but as it is an order of magnitude fainter than the first-reported companion in this system, we do not consider it in this work.}, showing compact cores and fainter, extended emission; \citet{2019ApJ...882...10N} argue that this extended structure is the result of gas being stripped in the early stages of a merger. Additional extended emission is seen in a rest-frame Ly$\alpha$ halo stretching north of the quasar, as reported by \citet{2019ApJ...881..131D} and \citet{2019ApJ...887..196F}.  From IR and [\ion{C}{2}] measurements, the companion is rapidly forming stars ($\textrm{SFR}\sim750\,\textrm{M}_\odot\,\textrm{yr}^{-1}$), yet it has an unobscured star formation rate of $\textrm{SFR}_\textrm{UV}<3\,\textrm{M}_\odot\,\textrm{yr}^{-1}$, implying that the companion is heavily obscured \citep{2019ApJ...881..163M}. 

The rest-frame UV spectrum of \shortname\ originally presented by \citet{2017ApJ...849...91M} is shown in Figure \ref{fig:nir_spec}. While the spectrum was analyzed in that work, the quasar itself was not classified. As broad absorption troughs are clearly present slightly blueshifted from the peaks of the \ion{Si}{4} and \ion{C}{4} lines, we classify \shortname\ as a broad absorption line (BAL) quasar. In particular, this quasar is a high-ionization BAL (HiBAL) quasar \citep{2000ApJ...538...72B}.
Based on the UV spectrum, \citet{2017ApJ...849...91M} report a black hole of $M_{\rm BH} = 3.05^{+0.44}_{-2.24}\times 10^9\ \textrm{M}_\odot$ that is accreting at an Eddington ratio of $L_{\rm bol}/L_{\rm Edd} = 0.48^{+0.11}_{-0.39}$. From ALMA observations, \citet{2019ApJ...882...10N} calculate a dynamical mass for the host galaxy of $\left(2.0 - 6.2 \right) \times 10^{10}\ \textrm{M}_\odot$ and $\left(2.7 - 8.4 \right) \times 10^{10}\ \textrm{M}_\odot$ for the companion. 
\begin{figure*}
\begin{center}
\includegraphics{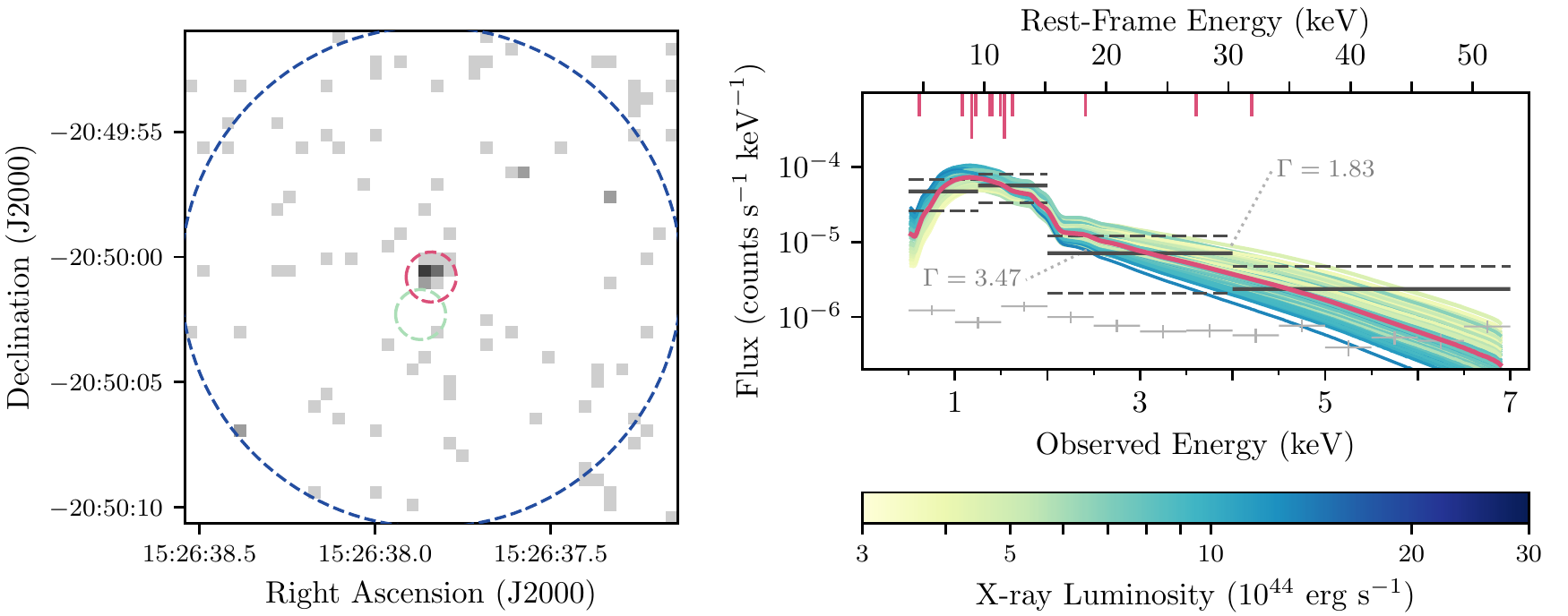} 
\end{center}
\caption{{\bf Left}: $0.5-7.0$ keV {\it Chandra} observation of \shortname. The quasar and companion are indicated by the red and green circles, respectively, while the large dashed blue circle traces the inner radius of the background annulus. While the quasar is clearly detected, the companion is not. {\bf Right}: X-ray spectrum of \shortname. Data (dark gray) are binned for ease of display, but were not binned during fitting. The best-fitting spectrum is shown in red, while 100 spectra  with $\Delta C \leq 2.30$ from our Monte Carlo analysis are shown colored by their X-ray luminosity. Higher values of luminosity are derived for softer power-law slopes; $\Gamma$ values for the two spectra at the extremes are marked. Energies of individual photons detected are indicated by vertical bars at the top of the figure, and the background flux level is shown in light gray in 0.5 keV wide bins. }\label{fig:sky_and_spectrum}
\end{figure*}

We adopt a Galactic neutral hydrogen column density of $N_{\rm H} = 8.35 \times 10^{20}\,\textrm{cm}^{-2}$ in the direction of \shortname\ \citep{2005A&A...440..775K}. While \citet{2019ApJ...882...77C} report a foreground absorption system at $z=6.476$, it is of low enough density ($ \log [N_{\rm C\ II} / \textrm{cm}^{-2} ] = 13.5$) that we do not consider its impact on our observations in this work. We use a flat $\Lambda$CDM cosmology with $H_0 = 70\,\textrm{km\,s}^{-1}\,\textrm{Mpc}^{-1}$, $\Omega_M = 0.3$, and $\Omega_\Lambda = 0.7$, and adopt a quasar redshift of $z=6.5864$, at which the scale is $5.42\,\textrm{pkpc}\,\textrm{arcsec}^{-1}$. Errors are reported at the 1$\sigma$ (68\%) confidence level unless otherwise stated. Upper limits correspond to $3\sigma$ limits.

\section{Observations}\label{sec:Observations}
\begin{deluxetable}{rcrr}
\tablecaption{Summary of {\it Chandra} Observations}
\label{tab:CXOobs}
\tablewidth{0pt}
\tablehead{
\colhead{Obs ID} & \colhead{Exposure Time} & \multicolumn2c{Start Time (UTC)}\\
\colhead{} & \colhead{(ks)} & \colhead{(YYYY-mm-dd)} & \colhead{(hh:mm:ss)}}
\startdata
20469 & 16.83 & 2019-05-27 & 08:00:46\\ 
22231 &	29.67 & 2019-05-31 & 01:44:11\\ 
22232 &	21.78 & 2019-05-31 & 19:42:49\\ 
22233 &	39.67 & 2019-06-01 & 14:36:29\\ 
22165 &	32.57 & 2019-06-20 & 05:36:12 
\enddata
\end{deluxetable}

We observed \shortname\ with the {\it Chandra X-ray Observatory} for $140.52$ ks across five separate visits. The details of these observations are given in Table \ref{tab:CXOobs}. All observations were conducted with the Advanced CCD Imaging Spectrometer \citep[ACIS;][]{2003SPIE.4851...28G}, with \shortname\ positioned to be observed with the back-illuminated ACIS S-3 chip. We used the Very Faint telemetry format and the Timed Exposure mode for our observations. 

Observations were reduced and analyzed with the {\it Chandra} Interactive Analysis of Observations software package \citep[CIAO,][]{2006SPIE.6270E..1VF} v4.11 using CALDB version 4.8.4.1. We first reduced our observations with the \texttt{chandra\_repro} script, using standard grade, status, and good time filters, as well as accounting for the telemetry format by setting \texttt{check\_vf\_pha=yes}. To minimize the impact of pointing uncertainties from the five observations, we used \texttt{WAVDETECT} \citep{2002ApJS..138..185F} and the CIAO tools \texttt{wcs\_align} and \texttt{wcs\_update} to align all five observations. For imaging analysis, the observations were then combined with the \texttt{merge\_obs} script into broad (0.5--7.0 keV), soft (0.5--2.0 keV), and hard (2.0--7.0 keV) images. We then ran the alignment procedure once more, this time using the full depth of our {\it Chandra} images to align with the Guide Star Catalog v2.3 \citep[GSC,][]{2008AJ....136..735L}. After correction, the absolute astrometry of our observations agreed with that of the GSC to an average of ${\sim}0\farcs7$, and the centroid position of \shortname\ agrees with the ALMA measurement \citep{2017Natur.545..457D} to less than a pixel ($0\farcs49$).

For both photometry and spectroscopy, we used a $1\farcs0$ radius aperture centered on the coordinates of \shortname, with a background in a concentric annular region with inner and outer radii of $10\farcs0$ and $30\farcs0$, respectively. A $1\farcs0$ radius aperture was also used to evaluate the companion. Both apertures and the inner radius of the background annulus are shown in Figure \ref{fig:sky_and_spectrum}. We extracted spectra from each individual observation using the task \texttt{spec\_extract} with the flag \texttt{correctpsf=yes}. Spectra were combined with the \texttt{combine\_spectra} task and analyzed with \texttt{XSPEC} v12.10.1 \citep{1996ASPC..101...17A} through the Python-based \texttt{PyXspec}.

\section{X-Ray Properties of \texorpdfstring{\shortname}{PJ231--20}} \label{sec:quasar}

In the broad band, we detect \shortname\ with $13.3^{+4.8}_{-3.7}$ net counts, computing source statistics with the method of \citet{1986ApJ...303..336G}. In the same photometric aperture, we detect $10.8^{+4.4}_{-3.3}$ and $2.5^{+2.9}_{-1.7}$ net counts in the soft and hard bands, respectively, up to a maximum detected energy of 4.20 keV. The hard band detection on its own would only be expected as a background fluctuation with probability $P=0.0118$, using binomial statistics \citep{2007ApJ...657.1026W}. From these photons, we compute the hardness ratio, $\mathcal{HR}$\footnote{$\mathcal{HR} = (H-S)/(H+S)$, where $H$ and $S$ are the net counts in the hard (2.0--7.0 keV) and soft (0.5--2.0 keV) bands, respectively.}, using the Bayesian methodology described by \citet{2006ApJ...652..610P}. Assuming uniform (Jeffreys) priors and integrating the posterior distribution with Gaussian quadrature, we find $\mathcal{HR} = -0.60_{-0.27}^{+0.17}$. In comparison, for PSO~J308.0416$-$21.2339, another $z>6$ quasar with a [\ion{C}{2}]-bright companion, \citet{2019ApJ...887..171C} report an observed $\mathcal{HR} = -0.48^{+0.11}_{-0.10}$ with a best-fit power-law index of $\Gamma=2.39^{+0.37}_{-0.36}$. 

To place further constraints on the properties of the quasar, we fit the X-ray spectrum in \texttt{XSPEC}. With so few counts, we do not bin the spectrum, and we use the modified C-Statistic \citep[$C$,][]{1979ApJ...228..939C, 1979ApJ...230..274W} to find the parameters of best fit.  As both the source and the background are Poisson data, the spectral fitting does not subtract a background, but instead accounts for both the model and the background simultaneously when evaluating the quasar spectrum (\texttt{XSPEC}'s W Statistic). We adopt the simple spectral model of \texttt{phabs}$\times$\texttt{powerlaw}, with $N_H$ frozen at its adopted value. The only parameters allowed to vary are the normalization of the power law and the spectral slope, $\Gamma$, so that $1\sigma$ uncertainties include all values where $\Delta C \leq 2.30$ (e.g., \citealt{1976ApJ...208..177L}; as noted by \citealt{1979ApJ...228..939C}, $\Delta C$ behaves as $\Delta \chi^2$ for evaluating confidence intervals).
\begin{figure*}
\begin{center}
\includegraphics{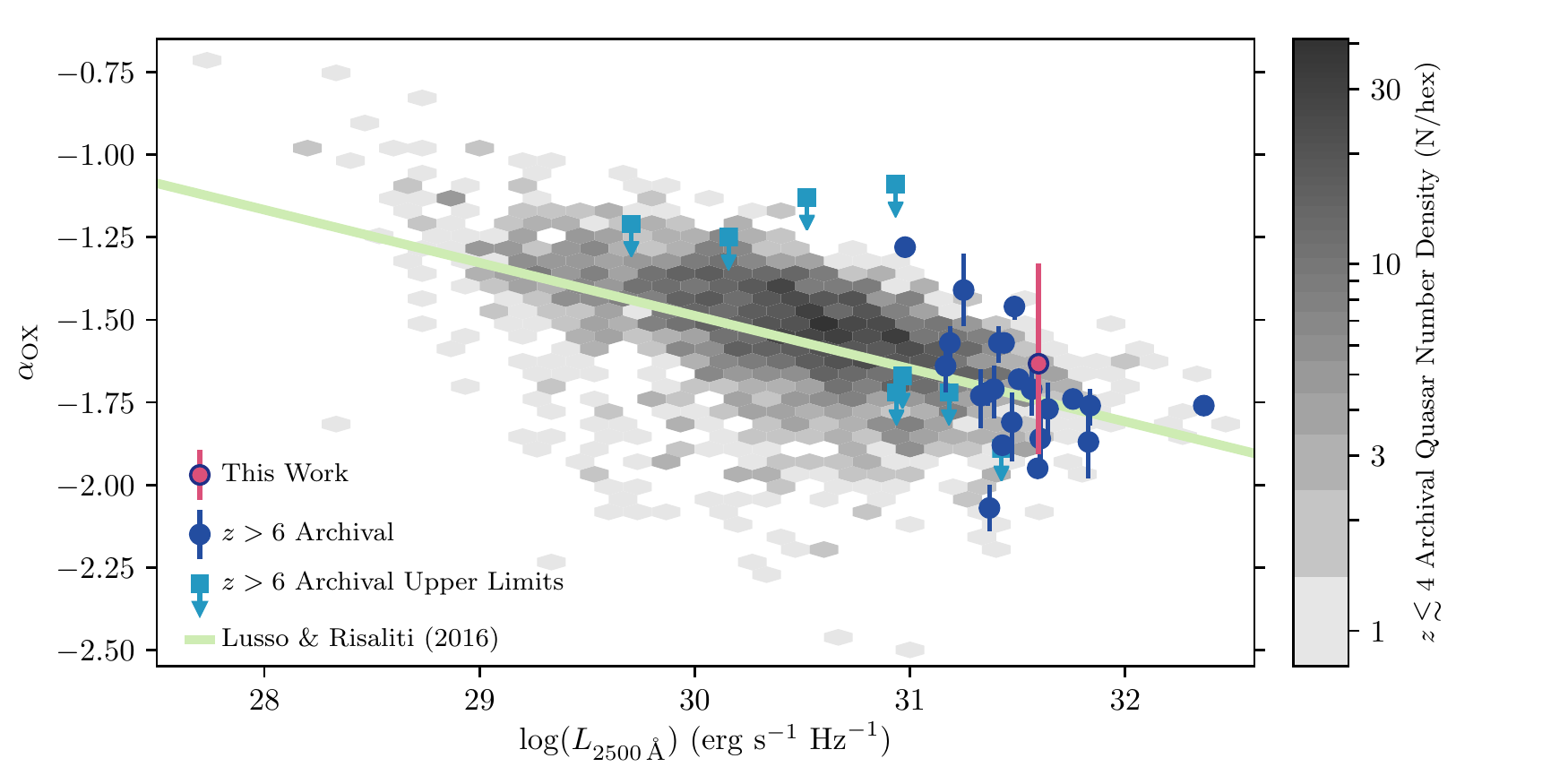}
\end{center}
\caption{Distribution of $\alpha_{\rm OX}$ as a function of rest-frame 2500 \AA\ luminosity for high redshift ($z>6$; points) and low redshift ($z \lesssim 4$, grayscale density plot) quasars, with \shortname\ highlighted. The underlying distribution and the best-fit to this trend (indicated by the green line) are both from \citet{2016ApJ...819..154L}, which is based on spectroscopically-confirmed broad-line quasars with {\it XMM} observations. The high redshift population is drawn from the compilation of \citet{2019A&A...630A.118V}, with later additions from \citet{2019ApJ...887..171C} and \citet{2020MNRAS.491.3884P}.
}\label{fig:a_ox}
\end{figure*}

We find the best fit for our spectrum with $\Gamma=2.6^{+1.0}_{-0.9}$, and a corresponding broad band flux of $F_{0.5-7.0} = 1.5^{+0.8}_{-0.5}\times 10^{-15}\ \textrm{erg}\ \textrm{s}^{-1}\ \textrm{cm}^{-2}$ and an unobscured luminosity in the rest frame 2.0--10.0 keV band of $L_{2-10} = 1.09^{+2.20}_{-0.70}\times 10^{45}\ \textrm{erg}\ \textrm{s}^{-1}$, where the uncertainties in  flux and luminosity include those of both fitted parameters. The results of our fit are summarized in Table \ref{tab:XProps}. The observed data, as well as 100 spectra from a Monte Carlo exploration of the parameter space, are shown in the right panel of Figure \ref{fig:sky_and_spectrum}.  The Markov Chain Monte Carlo method used to generate these spectra was the \texttt{XSPEC} routine \texttt{chain} using the \citet{2010CAMCS...5...65G} algorithm, with a burn length of 500 steps and a run length of 2500 steps. As shown in Figure \ref{fig:sky_and_spectrum}, the measured luminosity is being driven to higher values by an extrapolation of an extremely soft power-law index \citep[c.f.,][]{2019A&A...630A.118V} into the rest-frame soft-energy bands. If we fix $\Gamma$ to $\Gamma=2.2$ (as found by \citealt{2019A&A...630A.118V} to be representative of $z>6$ quasars) or $\Gamma=1.9$ \citep[typical of the quasar population in general, e.g.,][]{2017A&A...603A.128N}, we find best-fit luminosities of $L_{2-10} = 0.70\times 10^{45}\ \textrm{erg}\ \textrm{s}^{-1}$ and $L_{2-10} = 0.51\times 10^{45}\ \textrm{erg}\ \textrm{s}^{-1}$, respectively.

\begin{deluxetable}{rrr}
\tablecaption{X-ray Properties}
\label{tab:XProps}
\tablewidth{0pt}
\tablehead{
\colhead{Parameter} & \colhead{Value} & \colhead{Units}}
\startdata
Net Counts & $13.3^{+4.8}_{-3.7}$ & \nodata \\
Soft Counts & $10.8^{+4.4}_{-3.3}$ & \nodata \\
Hard Counts & $2.5^{+2.9}_{-1.7}$ & \nodata \\
$\mathcal{HR}$ & $-0.60_{-0.27}^{+0.17}$ & \nodata \\
$\Gamma$ & $2.6^{+1.0}_{-0.9}$ & \nodata\\
$L_{2-10} $ & $ 1.09^{+2.20}_{-0.70}\times 10^{45}$ & $\textrm{erg}\ \textrm{s}^{-1}$ \\
$F_{0.5-2.0} $ & $ 1.0^{+0.8}_{-0.4}\times 10^{-15}$ & $\textrm{erg}\ \textrm{s}^{-1}\ \textrm{cm}^{-2}$ \\
$F_{0.5-7.0} $ & $ 1.5^{+0.8}_{-0.5}\times 10^{-15}$ & $\textrm{erg}\ \textrm{s}^{-1}\ \textrm{cm}^{-2}$ \\
$\alpha_{\rm OX}$ & $-1.63^{+0.30}_{-0.27}$ & \nodata \\
$C / {\rm d.o.f}$ & 78.7 / 444 & \nodata
\enddata
\end{deluxetable}

Other analyses of high redshift quasars \citep{2017MNRAS.470.1587A, 2018A&A...614A.121N, 2019ApJ...887..171C} have also included the contributions from obscuration at the redshift of the quasar in their spectral fitting. With only 14 observed counts, adding a third free component to our model is ill-advised (and impractical). However, if we fix $\Gamma = 2.6$ and model the spectrum with the \texttt{XSPEC} model \texttt{zphabs}$\times$\texttt{phabs}$\times$\texttt{powerlaw}, again fixing the redshift and Galactic column density to their adopted values, we find that the $1\sigma$ uncertainties on the redshifted column density are $N_{H,z} \in [0.00, 4.8] \times 10^{23}\ \textrm{cm}^{-2}$. With such large uncertainties, we do not gain any meaningful insight by including this term; and, as \textit{Chandra} is insensitive to obscurations of $\lesssim 10^{23}\ \textrm{cm}^{-2}$ at these redshifts, ignoring \texttt{zphabs} does not affect our fitted results. We therefore do not include the \texttt{zphabs} term in the remainder of our analysis, although we do discuss the potential of large obscuration in Section \ref{sect:discussion}.

Lastly, we estimate the X-ray-to-optical power law slope, and compare the measured X-ray luminosity to that predicted from the brightness reported by \citet{2017ApJ...849...91M}. We characterize the X-ray-to-optical power law slope with $\alpha_{\rm OX},$ defined as  
\begin{equation}\label{eqn:aox}
\alpha_{\rm OX}= 0.3838 \times \log( {L}_{2\,\mathrm{keV}} / {L}_{2500\,\textup{\footnotesize \AA}}),
\end{equation}
where $L_{2\,\mathrm{keV}}$ and \luv\ are the monochromatic luminosities at rest-frame 2 keV and 2500 \AA, respectively. \citet{2017ApJ...849...91M} report a flux density of \fuv$ = 5.04_{-0.75}^{+0.03}\times 10^{-18}\ {\rm erg}\ {\rm s}^{-1}\ {\rm cm}^{-2}\ \text{\AA}^{-1}$ as measured from the NIR spectrum. Accounting for redshift k-corrections, this translates to a value of \luv$=3.95 \times 10^{31}\ {\rm erg}\ {\rm s}^{-1}\ {\rm Hz}^{-1}$.
From our data, we find $\alpha_{\rm OX}=-1.63^{+0.30}_{-0.27}$. For comparison, from the calculated value of \luv\ and from the observed relation of \citet{2016ApJ...819..154L}, we would expect $\alpha_{\rm OX}=-1.74^{+0.23}_{-0.23}$ (both sets of $\alpha_{\rm OX}$ errors include the uncertainties on \luv). The observed value of $\alpha_{\rm OX}$ is shown relative to that of low redshift and other high redshift quasars, as well as the fitted relation from \citet{2016ApJ...819..154L}, in Figure \ref{fig:a_ox}.

As a final note on our spectral fitting, we consider the effects of fitted bandpass and background region. All of our analyses were conducted in the traditional broad band (0.5--7.0 keV); however, as shown in Figure \ref{fig:sky_and_spectrum}, most of the observed flux is at soft energies. To see if including noisier energy ranges affected our results, we again fit the data, this time limiting ourselves to only 0.5--3.0 keV. However, we find that the results are not significantly changed, with a best-fit value of $\Gamma = 2.7^{+1.2}_{-1.3}$. We also tested the effect of varying the size of the region used for background extraction; systematic uncertainties produced by this selection are of order one-tenth the statistical uncertainties in our fit.

\section{Companion}

\shortname\ was targeted for observations with {\it Chandra} in part to search for X-ray emission from the companion detected by \citet{2017Natur.545..457D}. However, unlike PSO~J308.0416$-$21.2339 \citep{2019ApJ...887..171C}, where both the quasar and companion were seen, the companion here is not even detected at low significance. No photons are detected within $0\farcs9$ of the companion's position, and only one photon is observed within $1\farcs0$. This photon has energy 846 eV; the encircled energy fraction for {\it Chandra} sources at this energy exceeds 80\% within $0\farcs5$. In contrast, the probability of one photon arising in this aperture based on the measured background in the broad band is $P=0.51$, based on binomial statistics \citep{2007ApJ...657.1026W, 2014ApJ...785...17L}. As the probability of being a background fluctuation exceeds that of being a companion photon, we assume that we have detected no net flux from the companion itself. Here, we discuss what this non-detection means for both this companion and other [\ion{C}{2}]-detected companions in general.

First, we evaluate limits on the potential luminosity of the companion. Assuming Poisson statistics, the probability of zero photons being detected reaches $P=0.00135$ (i.e., a $3\sigma$ event) when the expected value is 6.61 photons. Using the \texttt{XSPEC} model \texttt{zphabs}$\times$\texttt{phabs}$\times$\texttt{powerlaw} and the response files of our observation, we calculate the unobscured luminosity of a model with 6.61 predicted counts. For a redshifted absorber of column density $N_{H,z} = 10^{23}\ \textrm{cm}^{-2}$, the limiting luminosity is $3.0 \times 10^{44}\ {\rm erg}\ {\rm s}^{-1}$ ($\Gamma = 2.0$) or $5.9 \times 10^{44}\ {\rm erg}\ {\rm s}^{-1}$ ($\Gamma = 2.6$). For a Compton-thick source ($N_{H,z} = \sigma_{\rm T}^{-1} = 1.5 \times 10^{24}\ \textrm{cm}^{-2}$), these luminosities become $6.3 \times 10^{44}\ {\rm erg}\ {\rm s}^{-1}$ ($\Gamma = 2.0$) and $15.2 \times 10^{44}\ {\rm erg}\ {\rm s}^{-1}$ ($\Gamma = 2.6$). Barring extreme levels of obscuration, we can therefore rule out an AGN in the companion with luminosity similar to that of the optically-selected quasar, despite both galaxies having similar masses.

A lack of observed emission from the companion is perhaps not surprising when considering the earlier work of \citet{2019ApJ...881..163M}. They reported that the companion is also not detected in rest-frame UV or optical emission, implying a significant amount of dust obscuration. Indeed, the spectral energy distribution of the companion as presented in that work is most like Arp 220, a well-known ultraluminous infrared galaxy in the local universe, being both highly star-forming and highly dust obscured. The X-ray luminosity of the AGN in Arp 220 is of order $L_{2-10}\lesssim10^{42}\ {\rm erg}\ {\rm s}^{-1}$ \citep{2017ApJ...841...44P}, and no AGN emission is seen in hard X-rays by {\it NuSTAR} \citep{2015ApJ...814...56T}. As Arp 220 has a SMBH with mass of order ${\sim}10^{9}\ {\rm M}_\odot$ \citep{2017ApJ...836...66S}, and as the ${\sim}100\ \mu{\rm m}$ rest-frame luminosities of Arp 220 and the companion only differ by a factor of ${\sim}1.4$ \citep{2019ApJ...881..163M}, it is thus possible to envision that the companion might only host a faint AGN.

Looking toward the broader population of [\ion{C}{2}]-detected companions \citep[six such systems are known at $z \gtrsim 6$;][]{2017Natur.545..457D, 2017ApJ...850..108W, 2019ApJ...882...10N}, the non-detection in X-rays of a companion around \shortname\ represents the first complete non-detection in three such systems well-studied so far (\citealt{2019A&A...628L...6V}; \citealt{2019ApJ...887..171C}). We note that SDSSJ0842+1218 has also been observed in X-rays \citep{2019A&A...630A.118V} and that no emission from its companion was seen; however, these observations were relatively short (29 ks), and so we treat the companion as being unobserved in this discussion. The two potentially detected companions (PSO J308--21 and PSO J167--13) are in systems far along in the merging process (with projected separations of ${\lesssim}10$ pkpc), as is one of the unobserved companion systems (SDSS J1306+0356). However, the other two remaining systems (SDSS J0842+1218 and CFHQ J2100--1715) are more like \shortname, with spectral energy distributions being dominated by star formation and dust obscuration \citep{2019ApJ...881..163M} and with star-forming companions that are still 10's of kpcs away from merging. \citet{2019A&A...623A.172C} recently put forward a model that, in the early universe, the dense interstellar medium of galaxies can cause obscuration supplementing or even surpassing that of the circumnuclear gas; in this context, AGN in companions might only be visible once a merger has begun to disrupt the gas distribution in the companion galaxy. It seems likely, therefore, that while a quasar may have a [\ion{C}{2}]-bright companion, that should not be taken as an indication of an X-ray-visible AGN in the companion, at least with current facilities.

\begin{figure*}
\begin{center}
\includegraphics{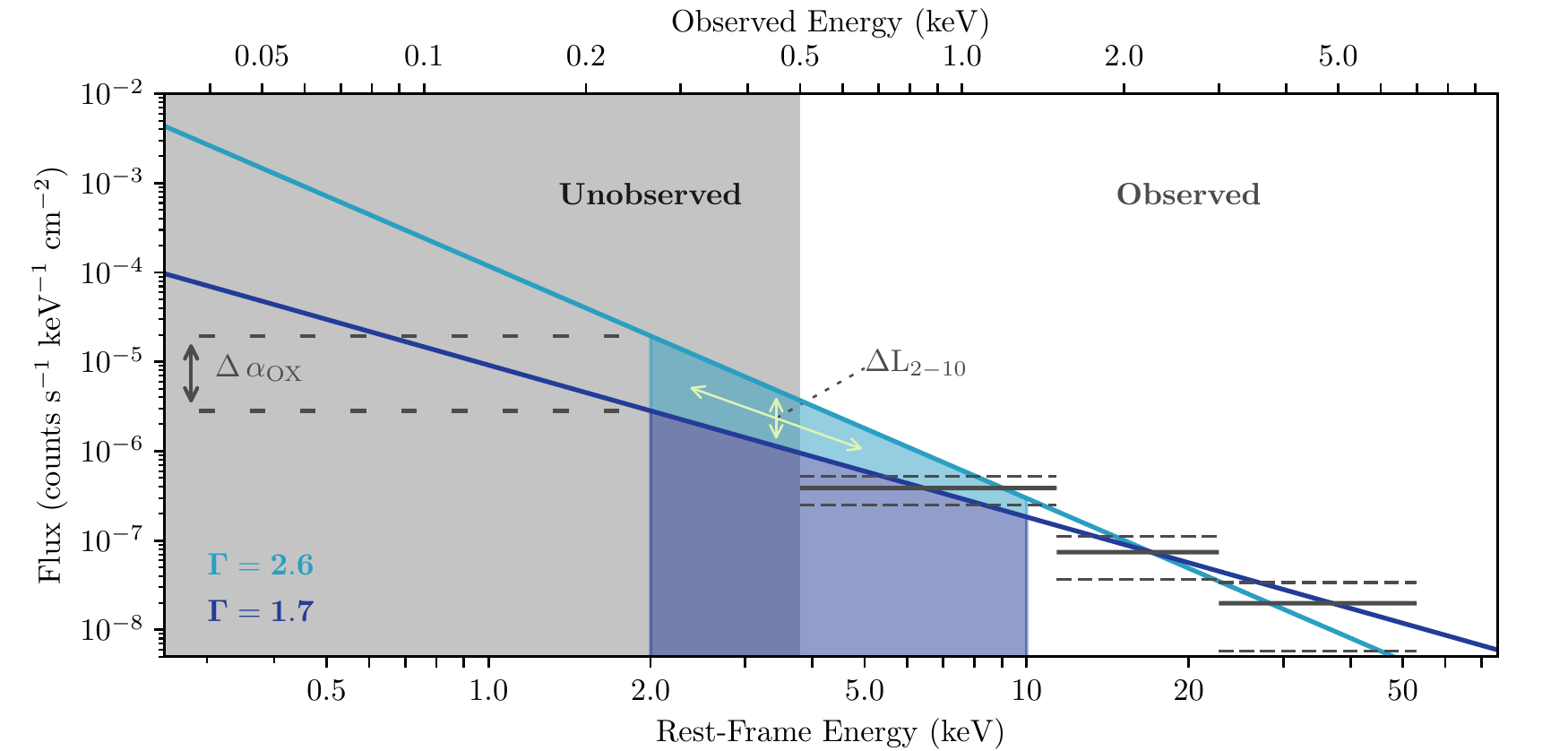}
\end{center}
\caption{Conceptual model of how interpreting \shortname\ as either soft (light blue) or faint (dark blue) affect calculated properties of the quasar. Our observed data (here corrected for the differing response of {\it Chandra} with energy) are shown binned by horizontal bars, with $1\sigma$ uncertainties as dashed lines above and below. The energy domain where we lack observations is indicated by the shaded gray region. Two power-law spectra -- which are toy models, not fits -- are shown; it is clear from this demonstration that both $\alpha_{\rm OX}$ and ${\rm L}_{2-10}$ are driven by extrapolations into the regime where we lack observations. }\label{fig:interpretations}
\end{figure*}
\vspace{7mm} 

\section{Interpretations} \label{sect:discussion}

With so few counts detected in the X-ray spectrum of \shortname\ despite the long exposure (cf. \citealt{2018ApJ...856L..25B}), it is difficult to definitively establish the properties of the AGN powering the quasar emission. Here, we discuss two possible interpretations to explain the data. First, we consider if the quasar spectrum is indeed as soft in the X-rays as given by the best fit, such that it still has an X-ray luminosity of the order $10^{45}\ {\rm erg}\ {\rm s}^{-1}$ despite its faint observed flux. Alternatively, we consider if the quasar spectrum is typical ($\Gamma \approx 2.0$), in which case there is not significant flux just below our soft-energy cutoff and the quasar is thus not highly luminous in X-rays. Finally, we discuss what effects obscuration and variability could have on our interpretations. 

\subsection{Soft X-ray Spectrum}

The first interpretation of these results is that the best fit is accurate -- the X-ray spectrum of \shortname\ is very soft, driving a fairly normal X-ray luminosity despite the small number of observed photons. In this context, the low count rate is a product of {\it Chandra} sampling far down the quasar's power-law spectrum due to the high redshift (a schematic of this interpretation is given in Figure \ref{fig:interpretations}). If this interpretation holds, and if \shortname\ is representative of some population of high redshift quasars, it spells trouble for deeper studies with currently-available facilities.

One potential reason for this spectral shape is that \shortname\ is undergoing a significant burst of accretion in connection with the ongoing merger. A number of works (e.g., \citealt{1995MNRAS.277L...5P}; \citealt{2003MNRAS.343..164B}; \citealt{2004A&A...422...85P}; \citealt{2013MNRAS.433.2485B}, \citeyear{2016ApJ...826...93B}; \citealt{2017MNRAS.470..800T}; \citealt{2018MNRAS.480.1819R}; \citealt{2019MNRAS.487.2463W}) have identified a trend between $\Gamma$ and $\lambda_{\rm Edd}$, where $\lambda_{\rm Edd} = L_{\rm bol}/L_{\rm Edd}$, such that larger values of $\Gamma$ trend with increasing Eddington ratios. \shortname\ is accreting at an Eddington ratio of $\lambda_{\rm Edd} = 0.48^{+0.11}_{-0.39}$ \citep{2017ApJ...849...91M}, although we note that the black hole mass -- and thus the Eddington ratio -- were derived using a relation with a 0.55 dex scatter on its zero point \citep{2006ApJ...641..689V}, and thus Eddington or even super-Eddington accretion is not incompatible with previous data. For reference, the best-fit $\Gamma$ predicted by the relation of \citet{2016ApJ...826...93B} for $\lambda_{\rm Edd} = 0.48$ is $\Gamma=2.25$, although due to the significant uncertainties in both $\lambda_{\rm Edd}$ and the scaling relation, this value is suggestive, not predictive.

Previous works have explained this correlation between $\Gamma$ and $\lambda_{\rm Edd}$ by arguing that an enhancement in accretion will lead to a stronger UV flux from the accretion disk, which in turn more effectively cools the electron corona \citep[e.g.,][]{2013MNRAS.433.2485B}. A cooler corona, in turn, produces a softer photon index, all other things being equal \citep[e.g.,][]{2018MNRAS.480.1819R}. While it may be tempting to consider the large Ly$\alpha$ halo around \shortname\ reported by \citet{2019ApJ...881..131D} and \citet{2019ApJ...887..196F} as a source of excess UV photons, the low luminosity ($L_{{\rm Ly}\alpha}\sim10^{44}\ {\rm erg}\ {\rm s}^{-1}$) and large spatial extent of the halo itself (${\sim}6.6$ sq. arcsec) make its UV contribution to the corona negligible. 

Our best-fit spectrum has $\Gamma=2.6^{+1.0}_{-0.9}$, which is softer than expected of low redshift AGN (albeit within $1\sigma$); in a fit of ${\sim}800$ bright, $z\lesssim0.4$ AGN, \citet{2017ApJS..233...17R} report a median $\Gamma\approx1.8$, with 90\% of the values falling between $1.3\lesssim\Gamma\lesssim2.2$. However, the observed value of $\Gamma$ for \shortname\ is not particularly extreme for high redshift quasars. In their analysis of $z>6$ AGN, \citet{2019A&A...630A.118V} report an average value of $\langle\Gamma\rangle = 2.20_{-0.34}^{+0.39}$ for sources with fewer than 30 detected counts and an average $\langle\Gamma\rangle =2.13^{+0.13}_{-0.13}$ for quasars with at least 30 counts. The only other clearly detected high redshift quasar with a [\ion{C}{2}]-bright companion,  PSO~J308.0416$-$21.2339, also has a high power-law index: $\Gamma = {2.39}_{-0.36}^{+0.37}$ \citep{2019ApJ...887..171C}. \citet{2018A&A...614A.121N} describe a change in $\Gamma$ for one particular $z{\sim}6.3$ quasar over a 13 year baseline, where early observations are best-fit by $\Gamma=2.37^{+0.16}_{-0.15}$ in a 2003 {\it XMM} observation and $\Gamma=1.81^{+0.18}_{-0.18}$ in a 2017 {\it Chandra} observation (we discuss the potential of variability below). From this, it seems as if \shortname\ having an intrinsically soft value of $\Gamma$ is plausible. 

\subsection{Typical X-ray Spectrum}

A second interpretation of the data is that \shortname\ is not a highly luminous AGN ($L_{2-10} < 10^{45}\ \textrm{erg}\ \textrm{s}^{-1}$, e.g., \citealt{2015MNRAS.453.1946G}; \citealt{2018AstL...44..500K}), and that the small number of detected counts is indicative of a quasar producing relatively few X-ray photons, both in and out of the observed X-ray bands (see Figure \ref{fig:interpretations}). In this context, $\Gamma$ is most likely toward the lower end of the $1\sigma$ distribution of our best-fit values, but is, in turn, less of an outlier from the overall distribution of AGN properties. Even for shallow power law slopes, small number statistics and {\it Chandra}'s soft-energy response combine for a spectrum dominated by soft photons when only few counts are present; this is true without needing to invoke an extreme value of $\Gamma$. Through that lens, our X-ray data are still in keeping with a harder, fainter source.

As discussed in Section \ref{sec:quasar}, a lower value of $\Gamma$ reduces the luminosity of \shortname\ to $L_{2-10} \lesssim 0.7 \times 10^{45}\ \textrm{erg}\ \textrm{s}^{-1}$. This luminosity is either around or slightly fainter than the knee of the AGN luminosity function for $z \gtrsim 4$, which various works have measured to be $L^{*}_{2-10} \approx 0.5 - 1.0 \times 10^{45}\ \textrm{erg}\ \textrm{s}^{-1}$ \citep{2014MNRAS.445.3557V, 2015MNRAS.453.1946G, 2018AstL...44..500K}.
Likewise, with an absolute magnitude of ${\rm M}_{1450}=-27.14$, \shortname\ does not have an extreme UV luminosity \citep{2017MNRAS.466.1160M, 2019MNRAS.488.1035K}.
As an otherwise normal quasar, the simple explanation for \shortname\ is that its true value of $\Gamma$ is closer to that of a typical quasar, which, even for higher redshifts, is shallower than that found by the best-fit.

We also note that, as shown in Figure \ref{fig:nir_spec} and discussed in Section \ref{sec:intro}, the NIR spectrum of \shortname\ shows evidence of being a BAL quasar. BALs are routinely observed to be X-ray faint for their rest-frame UV luminosity \citep[e.g.,][]{2009ApJ...692..758G}, as characterized by $\Delta\alpha_{\rm OX} = \alpha_{\rm OX,\ Observed} - \alpha_{\rm OX,\ Model}$, the difference between observed and predicted values of $\alpha_{\rm OX}$. While obscuration could be invoked for this correlation, studies with {\it NuSTAR} \citep[e.g.,][]{2014ApJ...794...70L} and with moderate-redshift AGN \citep[e.g.,][]{2014ApJ...786...58M} have shown that BAL spectra, when accounting for obscuration, still have extrapolated values of $\alpha_{\rm OX}$ lower than predicted by their \luv. Of course, even for the lowest value of $\alpha_{\rm OX}$ allowed by the $1\sigma$ uncertainties, \shortname\ is still within the scatter of the \citet{2016ApJ...819..154L} fit. In addition, the trend of large, negative values of  $\Delta\alpha_{\rm OX}$ with BALs is reduced for BALs with absorption lines of relatively low depth such as \shortname\ \citep{2018MNRAS.479.5335V}. Nevertheless, a lower value of $\alpha_{\rm OX}$ as would be caused by a typical value of $\Gamma$ is more in keeping with the rest-frame UV spectrum.

\subsection{Other Concerns}

One potential explanation for an observed X-ray faintness is heavy dust obscuration of the AGN, but this does not mesh with the observed data. The observed hardness ratio of \shortname, $\mathcal{HR} = -0.60_{-0.27}^{+0.17}$ (see also Figure \ref{fig:sky_and_spectrum}), implies that most of the observed flux is at relatively soft energies. If a significant screen of dust existed, we would expect it to drive the hardness ratio to much harder values. Likewise, the broad \ion{Mg}{2}\ emission (${\rm FWHM} \sim 4700\ {\rm km}\ {\rm s}^{-1}$) reported by \citet[][see also Figure \ref{fig:nir_spec}]{2017ApJ...849...91M} indicates that \shortname\ is a Type 1 AGN; Type 1 AGN have long been seen to have minimal amounts of dust ($N_{H} \lesssim 10^{22}\ {\rm cm}^{-2}$; e.g., \citealt{2018ApJ...856..154S}). Due to the extreme redshift of \shortname, such column densities would have no impact on our observed spectrum. Thus, we do not expect dust to be responsible for the observed faintness of \shortname.

Previous optical and near-infrared observations of the \shortname\ system, as reported by \citet{2017ApJ...849...91M}, \citet{2019ApJ...881..163M}, \citet{2019ApJ...881..131D}, and \citet{2019ApJ...887..196F}, were conducted between March 2015 and July 2017. The X-ray observations described here, however, took place almost two years afterward (see Table \ref{tab:CXOobs}). It is therefore possible that the physical conditions of the AGN changed during that time -- although we note that a temporal offset of $\Delta t$ in the observed frame corresponds to an offset of $\Delta t / (1 +z)$ in the rest-frame of the AGN. Previously, \citet{2014MNRAS.440L..91P} reported an X-ray luminosity drop of ${\sim}2\times$ over 15 months (observed) for a $z\sim7.1$ quasar, and \citet{2018A&A...614A.121N} report variability over observed timescales of 16 months (low significance) and 14 years for one $z=6.31$ quasar, where total fluctuations were again on the order of $2\times$. It is thus unlikely that the observed X-ray photon faintness was caused by variability alone.

\section{Summary}
We have presented deep (140 ks) {\it Chandra} observations of \shortname, a $z=6.59$ quasar with a nearby (8 pkpc) companion detected in [\ion{C}{2}] with ALMA. Our primary results are:
\begin{itemize}
    \item \shortname\ is detected in the X-ray observations, with $13.3^{+4.8}_{-3.7}$ counts. The source is very soft, having a hardness ratio of $\mathcal{HR} = -0.60_{-0.27}^{+0.17}$. While the small number of detected counts makes the results of spectral fitting uncertain, a simple power-law model best fits the data with $\Gamma=2.6^{+1.0}_{-0.9}$.
    \item Despite the faint detection, the rest frame 2.0--10.0 keV luminosity is reasonable for the SMBH mass powering the \shortname\ quasar: $L_{2-10} = 1.09^{+2.20}_{-0.70}\times 10^{45}\ \textrm{erg}\ \textrm{s}^{-1}$. However, the luminosity is heightened by the large value of $\Gamma$; if $\Gamma$ is fixed to 2.2, more typical of high-redshift quasars, the best-fit luminosity drops to $L_{2-10} = 0.70\times 10^{45}\ \textrm{erg}\ \textrm{s}^{-1}$.
    \item The companion galaxy is not detected in X-rays. While at least one other companion galaxy has been detected by {\it Chandra} around a different $z>6$ quasar \citep{2019ApJ...887..171C}, previous {\it HST} observations revealed that the companion  to \shortname\ is heavily obscured \citep{2019ApJ...881..163M}. If an AGN is present, it is too faint and/or obscured to be visible to current X-ray facilities.
    \item We discussed two potential interpretations of the observed properties of \shortname. First, that it has an extremely soft power-law index, and therefore most of the emission is below that observable by {\it Chandra}. Second, that $\Gamma$ is actually smaller, and that the quasar is not highly luminous in the rest-frame soft X-ray band. Neither variability nor obscuration by themselves appear to play a significant role in explaining the small number of observed photons.
    
\end{itemize}

To better understand the conditions of the AGN of \shortname, further observations are required -- specifically, deeper X-ray observations. {\it XMM-Newton}, which has a larger effective area than {\it Chandra} that extends to softer energies (0.2 keV), may even allow for a characterization of \shortname's AGN with only a modest exposure time.
However, current facilities are being pushed to their limits to observe high-redshift quasars, and deeper understanding will only come with the launch of {\it Athena} \citep{2013arXiv1306.2307N} and, hopefully, {\it Lynx} \citep{2019JATIS...5b1001G} and {\it AXIS} \citep{2019BAAS...51g.107M}. 

\vspace{7mm} 
\acknowledgments

{\small The work of TC and DS was carried out at the Jet Propulsion Laboratory, California Institute of Technology, under a contract with NASA. TC's research was supported by an appointment to the NASA Postdoctoral Program at the Jet Propulsion Laboratory, California Institute of Technology, administered by Universities Space Research Association under contract with NASA. The scientific results reported in this article are based on observations made by the \textit{Chandra X-ray Observatory}. This research has made use of software provided by the \textit{Chandra} X-ray Center (CXC) in the application package CIAO. An additional 1920 kg of ${\rm CO}_2$ emissions were generated by TC on work-related travel during the preparation of this work, following the calculations of the \href{https://www.icao.int/environmental-protection/Carbonoffset/}{International Civil Aviation Organization}. 
}

\facility{CXO}
\software{BEHR \citep{2006ApJ...652..610P},
          CIAO \citep{2006SPIE.6270E..1VF},
          PyFITS \citep{1999ASPC..172..483B},
          WAVDETECT \citep{2002ApJS..138..185F},
          XSPEC \citep{1996ASPC..101...17A}}
          
\textcopyright\ 2020. All rights reserved.

\bibliography{bibliography}

\end{document}